# Magnetization reversal mechanism of double-helix nanowires probed by dark-field magneto-optical Kerr effect

T. Gokita,[1,a] J. Jurczyk,[1] N. Leo,[1,2] S. Koraltan,[1] A. Anadón,[1,3] M.A. Cascales-Sandoval,[1] R. Belkhou,[4] C. Abert,[5,6] D. Suess,[5,6] C. Donnelly,[7,8] and A. Fernández-Pacheco[1,a]

AFFILIATIONS
[1]Institute of Applied Physics, TU Wien, 1040 Vienna, Austria
[2]Department of Physics, School of Science, University of Loughborough, Loughborough LE113TU, United Kingdom
[3]Instituto de Nanociencia y Materiales de Aragón, CSIC-Universidad de Zaragoza, 50009 Zaragoza, Spain
[4]Synchrotron SOLEIL Saint-Aubin 91190, France
[5]Physics of Functional Materials, Faculty of Physics, University of Vienna, 1090 Vienna, Austria
[6]Research Platform MMM Mathematics-Magnetism-Materials, University of Vienna, 1090 Vienna, Austria
[7]Max Planck Institute for Chemical Physics of Solids, 01187 Dresden, Germany
[8]International Institute for Sustainability with Knotted Chiral Meta Matter (WPI-SKCM$^2$), Hiroshima University, Higashi-Hiroshima 739-8526, Japan
[a)]E-mail: takeaki.gokita@tuwien.ac.at and amalio.fernandez-pacheco@tuwien.ac.at

**ABSTRACT**
Double-helix (DH) nanowires provide a platform to study the influence of geometric chirality on spin chirality. Their three-dimensional (3D) helical architecture and tunable inter-strand coupling enable control of spin chirality, including the stabilization of topological 3D magnetic states such as helical domains and domain walls, topological stray fields, and extended helical vortex/skyrmion tubes. So far, the study of these and other 3D nanostructures is usually confined to a limited number of magnetic microscopy experiments in large facilities. Here, we investigate the reversal mechanism of a single DH nanowire using Dark-Field magneto-optical Kerr effect (DF-MOKE) magnetometry under external 3D magnetic fields. By analyzing the angular dependence of the DF-MOKE signal, we fit the reversal process using established models for domain-wall nucleation and propagation, finding a characteristic behavior similar to that reported for cylindrical nanowires. Micromagnetic simulations indicate that the reversal process goes through nucleation of the helical vortex tube in a curling manner while ptychographic X-ray magnetic circular dichroism data reveal that this helical vortex tube state forms through a mixed nucleation-propagation process. These observations provide a consistent microscopic picture of reversal mediated by a helical vortex tube extending along the nanowire. Our work provides a comprehensive characterization of magnetization reversal in DH nanowires and demonstrates that DF-MOKE magnetometry is effective for probing reversal mechanisms in single 3D nanostructures. This lab-based approach expands the range of accessible experiments beyond large-scale facilities, enabling extensive exploration of the rich spin states supported by 3D nano-geometries.

Three-dimensional (3D) nanomagnetism explores magnetic phenomena in three dimensions, offering new functionalities for spintronic devices and providing a platform for investigating novel physical effects.[1,2] In particular, the geometry of 3D nanostructures is increasingly attracting interest for controlling magnetic chirality. Magnetic chirality plays a vital role in magnetism and spintronics, stabilizing chiral spin textures,[3–5] facilitating fast DW motion,[5,6] and enabling non-reciprocal phenomena.[7,8] Specifically, we have been investigating the magnetic states in double-helix (DH) nanowires, which consist of two magnetic helical strands positioned very close together or in direct contact. In these systems, the tunability of the helical pitch and the inter-strand separation enables geometric control of the system's spin chirality and facilitates the stabilization of a variety of nontrivial magnetic states, including helical domains and domain walls (DWs),[9] topological stray field textures,[10] and vortex/skyrmion tubes.[11]

Studies of 3D magnetic systems such as these normally rely on imaging experiments performed at synchrotron or large electron microscopy facilities, which restrict accessibility and flexibility. As a result, only a limited number of measurements can typically be performed, making systematic investigations and the exploration of phase diagrams involving different spin states[10,11] particularly challenging. We recently demonstrated a new lab-based characterization technique for 3D magnetic nanostructures, the Dark-Field magneto-optical Kerr effect (DF-MOKE).[12,13] This method measures the MOKE signal from 3D nanostructures under vector magnetic fields by analyzing the Kerr response at the specific angle at which the light is specularly reflected from a 3D nanostructure, rather than from the substrate. DF-MOKE has been previously applied to single 3D ramped Permalloy nanostrips, either interconnected or decoupled from the substrate, where it demonstrated single-nanostructure sensitivity[13] and enabled controlled 3D DW injection experiments[12] Here, we extend the DF-MOKE technique to more complex 3D geometries, showing that it can be used to detect the magnetic reversal of DH nanowires. Additionally, by employing vector magnetic fields, we do not only detect the switching of 3D nanostructures but also probe their magnetization reversal mechanism. These measurements are complemented with micromagnetic simulations and synchrotron-based experiments, providing a comprehensive view of the spin configurations throughout the reversal process.

We fabricated cobalt 3D chiral nanowires on a thermally oxidized silicon substrate using focused electron-beam-induced deposition (FEBID), as illustrated in Fig. 1(a). The growth was controlled using the f3ast[14] software developed in our group for the fabrication of complex 3D nanostructures. We utilized $Co_2(CO)_8$ precursor gas, which yields highly pure cobalt.[15] The accelerating voltage during growth was 5 kV, and the beam current was 43 pA, with an initial growth rate of 120 nm/s. During fabrication, the substrate was tilted at 52° from the horizontal axis to facilitate nanowire growth at a low angle with the substrate, and the deposited structure has right-handed (RH) chirality, as shown in Fig. 1(b). The total length of the structure is 3.5 μm, the diameter of each strand is 85 nm, and the overlapping distance of the two strands is 18 nm as per the CAD design.

In this letter, we aim to determine the magnetization reversal mechanism of DH nanowires using a lab-based technique. For this purpose, we measured the magnetization switching field ($H_{SW}$) via DF-MOKE measurements as a function of angle $\theta$, which is defined as the relative angle between the external magnetic field direction and the principal axis of the nanostructure, as illustrated in Fig. 2(a). The experimental data of $H_{SW}$ has been typically described by three different analytical models: Stoner-Wohlfarth[16] (S-W), Kondorsky[17,18], and Curling[19–22]. In the S-W model, magnetization switching is considered to be a coherent rotation.[16] In this case, the angular variation of the switching field is expressed by[16]

$$H_{SW} = H_k/(\sin^{\frac{2}{3}}\theta + \cos^{\frac{2}{3}}\theta)^{3/2}, \qquad (1)$$

where $H_k$ is the anisotropy field. In the Kondorsky model, the switching process is dominated by the propagation of magnetic DWs,[17] and the angular variation of the switching field is expressed by[17,18]

$$H_{SW} = H_{SW}(\theta = 0°)/\cos\theta. \qquad (2)$$

The Curling model describes switching through a quasi-uniform rotation of the magnetization, characterized by a curling-type rotation,[20–22] and the angular variation of the switching field is expressed by[20–22]

$$H_{SW} = M_s a(1+a)/2\sqrt{a^2 + (1+2a)\cos^2\theta}, \qquad (3)$$

where $a = -1.08(d_0/d)^2$, $d$ is the diameter of the structure, and $d_0 = 2(A/M_s^2)^{1/2} \approx 5 l_{ex}$,[23] with $A$ the exchange stiffness, $M_s$ the saturation magnetization, and $l_{ex}$ the exchange length. Magnetic fields can be applied along any direction of space thanks to a custom-made hexapole electromagnet with optical access.[24]

Fig. 2(b) shows the hysteresis loop of the DH with the magnetic field applied parallel to the structure ($\theta = 0°$), averaging 60 separate DF-MOKE measurements. Normalized DF-MOKE loops taken at different field angles are summarized in Fig. 2(c), with colors indicating the applied field angles. Here, even after subtraction of the Faraday contribution, the gradual slopes remain uneven. This behavior may arise from magnetization components not aligned with the structure,[25] or from quadratic contributions to the signal[26] due to slight misalignment of structure and field direction. Despite this, a sharp switching of the nanowire remains visible for the whole angular range. $H_{SW}$ was determined from the magnetic field corresponding to the highest peak of the first derivative of the signal. Fig. 2(d) shows the $\theta$ dependence of the $H_{SW}$ (black). In our case, both the Kondorsky (blue) and Curling (red) models accurately reproduce the experimental results with Eq. (2) and (3), respectively, whereas the S.-W. (green) model fails to do so with Eq. (1). Such trends are commonly observed in cylindrical nanowires,[25,27] and applies also to our case, since the DH nanowire can be considered to be a cylinder with a corrugated surface. Treating the corrugation as an effective roughness, its amplitude is 6.7 nm, calculated as the standard deviation of the radius of the structure extracted from SEM images (see SI1 for details of analysis ). Because this roughness is much smaller than the probe wavelength (658 nm), the surface can be considered as an effective optically smooth medium, *i.e.*, specular reflection dominates and roughness-induced depolarization is negligible. Therefore, the standard Fresnel-based magneto-optic formalism can be applied to analyze the Kerr response, and the approach followed by DF-MOKE remains valid. From the Curling fit, we calculated the exchange length, $l_{ex} = 8 \pm 1$ nm, which shows a reasonably good agreement with the reported value of Co.[28] It should be noted that the Kondorsky and Curling models exhibit similar angular dependence, making it impossible to distinguish the dominant magnetization switching mechanism only from these measurements.

To gain additional information and qualitatively evaluate our experimental results, we performed micromagnetic simulations. For this purpose, we utilized the magnum.pi software, based on finite elements.[29] Fig. 3(a) illustrates the geometry for the simulations, where an external magnetic field is applied at different angles($\theta$) . The structure used in the simulations has the same geometry as the RH DH nanowire as in the experiments, except for a reduced length of 1.3 μm to accelerate computation. The magnetic parameters were set to a saturation magnetization, $M_s = 700$ kA/m, an exchange stiffness constant, $A_{ex} = 1.0 \times 10^{-11}$ J/m, and a magnetic damping constant, $\alpha = 1$ , typically employed to study relaxed states in Co grown by FEBID.[10] Fig. 3(b) shows the z'-component of magnetization as a function of the external magnetic field at different $\theta$ values, the same protocol as in experiments. The hysteresis loops broaden as $\theta$ increases, with the $\theta$ dependence of $H_{SW}$ shown in Fig. 3(c). Three different fittings are included to $H_{SW}$ extracted from simulations: S.-W. (green), Kondorsky (blue), and Curling (red). Consistent with the experimental results, $H_{SW}$ is well reproduced by both the Kondorsky and Curling models, indicating that micromagnetic simulations qualitatively reproduce the experimental results.

To understand the underlying spin configurations during reversal, the simulated magnetization configurations were examined in detail. As the field decreases from saturation, the magnetization first undergoes a quasi-uniform rotation, with spins tilting coherently toward the field direction.

Upon further reduction, the system minimizes its magnetostatic energy by reducing surface magnetic charges through an azimuthal curling of the magnetization, leading to the nucleation of a helical vortex tube that propagates along the nanowire (Fig. 3(d)–3(f)). The vortex tube itself exhibits a helical character, with its magnetic chirality dictated by the geometric chirality of the double-helix structure, as previously reported.[9,11] This evolution reflects a curling-type reversal mechanism, in which the transition from an almost uniform to a vortex-like state allows the system to lower its total energy while reversing its magnetization.

To experimentally validate and complement the DF-MOKE measurements and micromagnetic simulations, we conducted X-ray ptychography experiments, which enable us to obtain high spatial resolution images,[30–34] and investigated the evolution of spin states with external fields through X-ray magnetic circular dichroism (XMCD). For the synchrotron experiments, we fabricated the DH nanowires on a pre-cut copper Transmission Electron Microscopy grid using FEBID, with a 43 pA beam, and an initial growth rate of 70 nm/s. The chirality of the structure presented here is left-handed (LH) rather than the RH chirality of the sample investigated by MOKE. Because the chirality is uniform across each structure, we expect the magnetization reversal under this type of field sequence to be equivalent in both cases. The structure was slightly tilted ($\approx 30°$) from the vertical direction toward the X-ray beam, as shown in Fig. 4(a), allowing us to measure both the in-plane (x-axis $\parallel k$) and axial (z-axis $\perp k$) components, as the XMCD signal arises from the scalar product of the magnetization and the X-ray beam. We measured the hysteresis loop with an axial field ($B_z$) to see the evolution of the spin states. To obtain the XMCD signals, the photon energy was tuned to 778.75 eV (slightly below the Co $L_3$ absorption edge), and the corresponding reconstructed phase images were analyzed.[31,32] This off-edge approach reduces the absorption while providing an excellent magnetic contrast.[34]

Fig. 4(b) shows selected XMCD images taken while sweeping the field from negative to positive saturation (see SI2 for details of the XMCD analysis). The initial image shown in this sequence corresponds to $B_z = 15$ mT, where a uniformly magnetized state is observed, indicating full remanence. At 17 mT, a characteristic XMCD vortex tube contrast (white on the left and black on the right side) appears along half of the structure. As the field increases to 18 mT, this vortex region expands, indicating that the magnetization reversal proceeds through this vortex-mediated state. At 22 and 23 mT, the magnetization is fully reversed from negative to positive.

Fig. 4(c) presents the average XMCD signal extracted from the images, showing how the magnetization reversal evolves as a function of the external field for this branch of the hysteresis loop. The resulting hysteresis loop presents a double switching, an effect less evident in the DF-MOKE data or simulations. This suggests that, although the nucleation mechanism is curling-type, proceeding via the formation of a helical vortex tube, the subsequent propagation can be locally hindered by defects, leading to a secondary switching event. To compare the experimental results with numerically predicted states, we computed XMCD images from the micromagnetic simulations shown before. For this, the structure was tilted 30° from the vertical $z$ axis, and the field was applied parallel to it, as illustrated in Fig. 4(d). Fig. 4(e) shows the calculated XMCD images while sweeping the field from -100 mT to +100 mT. At -100 mT the magnetic state is fully saturated. The first image displayed corresponds to +23 mT, where the nanowire remains uniformly magnetized. Although the XMCD contrast appears uniform, a slight asymmetry between the left and right strands is visible, arising from their different projection angles of their spins.[11] A helical vortex tube emerges at +30 mT and evolves towards a modified configuration at +33 mT, where the XMCD contrast becomes nonuniform due to local deviations of the magnetization and partial reversal at the wire edges. At +34 mT the magnetization switches, and full saturation is recovered by +100 mT. Overall, the calculated XMCD images match the experimental ones well, confirming full remanence for this geometry and revealing that reversal is mediated by the nucleation and evolution of a vortex tube along the nanowire. The X-ray results therefore confirm the simulated evolution from a saturated state to the nucleation of a helical vortex tube, followed by its propagation and eventual annihilation as the magnetization reverses. These observations bridge the macroscopic DF-MOKE signatures with the microscopic spin textures predicted numerically, establishing a consistent picture of the reversal process in DH nanowires.

In summary, we investigated the magnetization reversal mechanism of single DH nanowires under fully controlled 3D magnetic fields. Angular-dependent DF-MOKE measurements provided a clear and robust signal for DH nanowires, extending the range of 3D nanostructures accessible to this technique and enabling direct comparison with analytical models. The magnetization reversal mechanism is similar to that of cylindrical nanowires, but is further influenced by the geometric chirality of the DH structure, which determines the magnetic chirality of the system. Micromagnetic simulations and ptychographic XMCD measurements reveal that reversal proceeds through the nucleation of a helical vortex tube followed by its propagation, providing the microscopic picture underlying the angular dependence of the magnetization switching detected by DF-MOKE measurements. This study establishes DF-MOKE magnetometry under 3D field control as a powerful, lab-based approach for systematic angular studies of magnetization reversal in individual 3D nanostructures. The methodology can be readily applied to nanostructures fabricated not only by FEBID, but also by multiphoton lithography[35] or other 3D nanofabrication techniques[36]. Moreover, the same field angle-resolved framework could be in principle extended to other optical methods, such as focused Brillouin light scattering (BLS). The addition of such non-disruptive, laboratory-based methods provides a versatile platform for probing and mapping the rich spin textures that emerge in magnetic nanostructures with complex 3D geometries.

This work was supported by the European Community under the Horizon 2020 program, Contract No. 101001290 (3DNANOMAG). C.D. was also supported by the Max Planck Society Lise Meitner Excellence Program and funding from the European Research Council (ERC) under the ERC Starting Grant No. 101116043 (3DNANOQUANT). The authors acknowledge financial support from the Blue-Sky Research Fund – Phoenix, TU Wien.

**DATA AVAILABILITY**

The data that support the findings of this study are available in reposiTUM (DOI to be included upon acceptance).

**Conflict of Interest**

The authors have no conflicts to disclose.

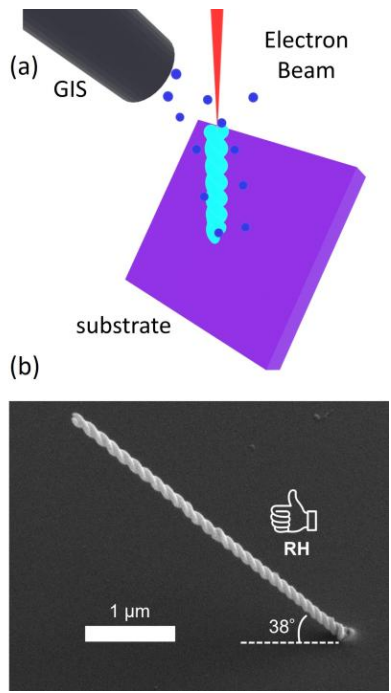

Fig. 1. (a) Schematic illustration of the FEBID process for the fabrication of DH nanowires made of cobalt. During the process, the substrate was kept at a 52° tilt. (b) SEM image of the deposited 3D cobalt right-handed DH nanowires, 38° tilted from the substrate (image was taken at 45° stage tilt).

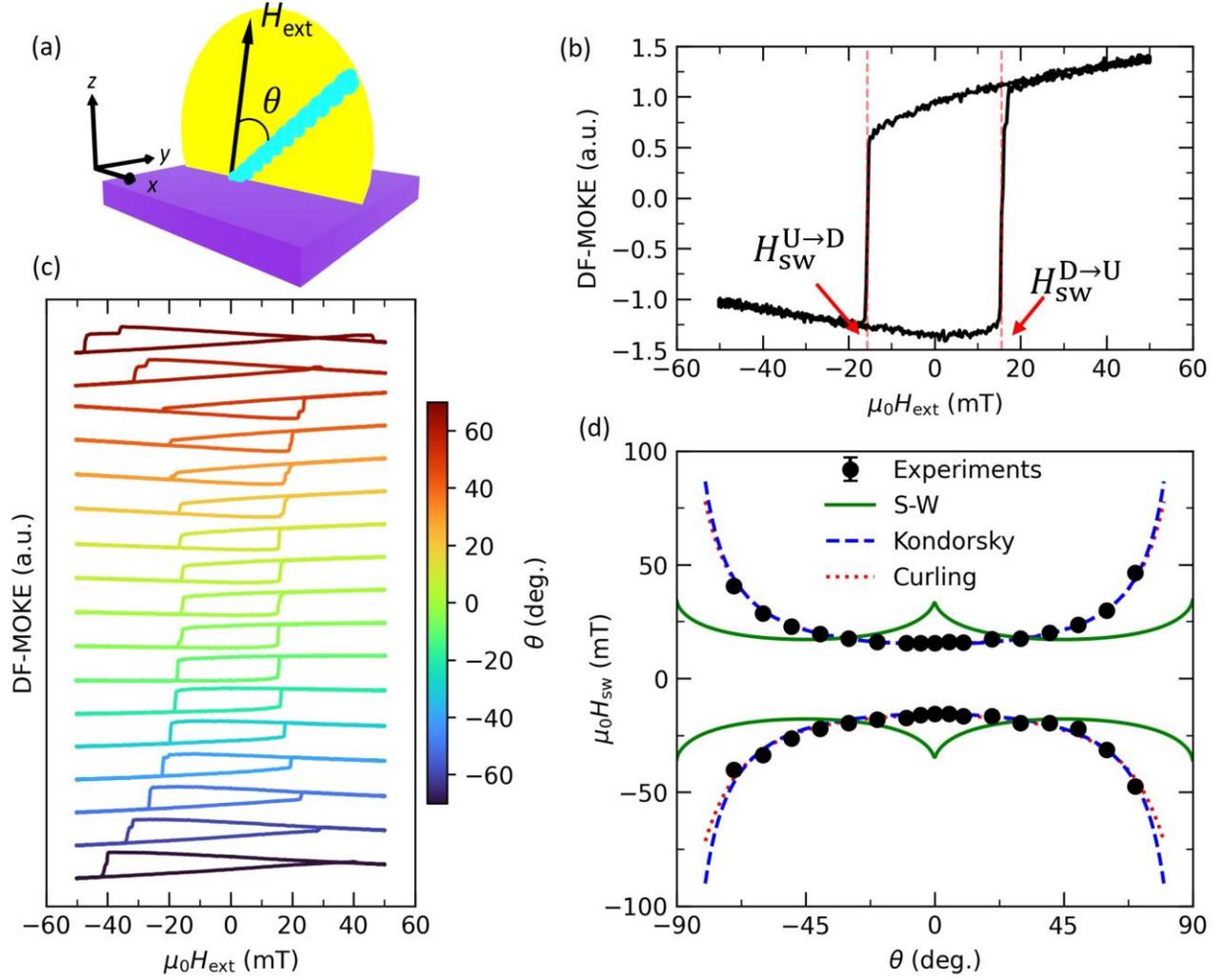

Fig. 2. (a) Schematic illustration of the relationship between the external magnetic field ($H_{ext}$) and the structure in angular-dependent DF-MOKE measurements. The angle $\theta$ is defined as the relative angle of $H_{ext}$ with respect to the nanowire. (b) DF-MOKE signal of a 3D nanowire as a function of the external magnetic field applied parallel to the nanowire ($\theta = 0°$). The red dashed line indicates the magnetization switching field ($H_{SW}$), calculated by the highest peak of the first derivative of the MOKE signal. $H_{SW}^{D \to U}$ denotes the switching field when the external magnetic field was swept from negative to positive, while $H_{SW}^{U \to D}$ corresponds to the switching field for the opposite sweep direction. (c) DF-MOKE loops measured at different $\theta$, with colors indicating the field direction. Data are offset for clarity. (d) Angular dependence of the switching field and three different fitting lines: Stoner-Wohlfarth (S-W) model (green), Kondorsky model (blue), and Curling model (red). The Kondorsky and Curling models provide a good fit (see text for details).

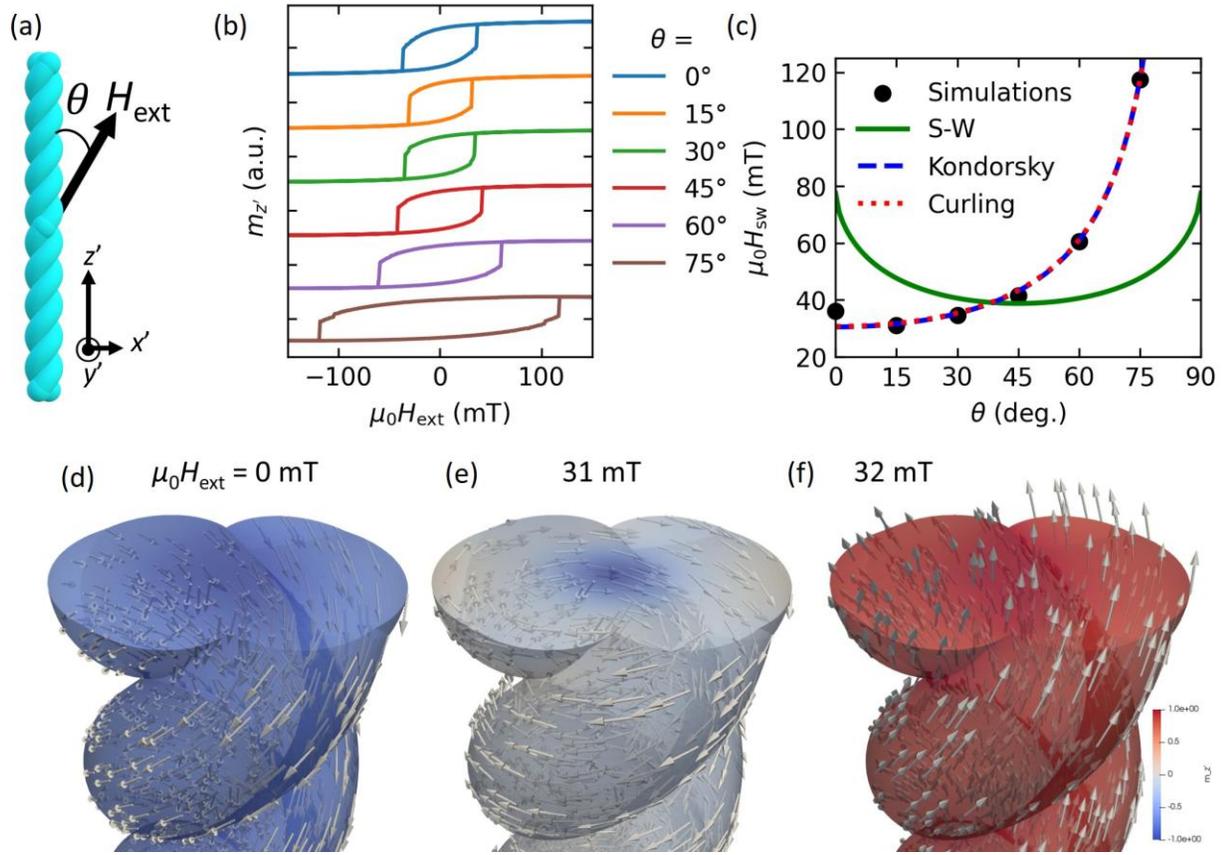

Fig. 3. (a) Schematic illustration of the geometry for the micromagnetic simulations. $\theta$ is defined as a relative angle of the external magnetic field ($H_{ext}$) with respect to the nanowire. (b) Magnetization curves ($z'$ component) as a function of $H_{ext}$ at several $\theta$. (c) $H_{SW}$ as a function of the $\theta$ and three different fitting lines: Stoner-Wohlfarth model (green), Kondorsky model (blue), and Curling model (red). Magnetic states snapshots obtained while sweeping the external magnetic field from -100 mT to +100 mT at $\theta$ = 15°. (d) Remanent state (0 mT), (e) helical vortex tube state before magnetization reversal (31 mT), and (f) state after reversal (32mT). Colors denote the normalized $z'$ component of the magnetization.
7

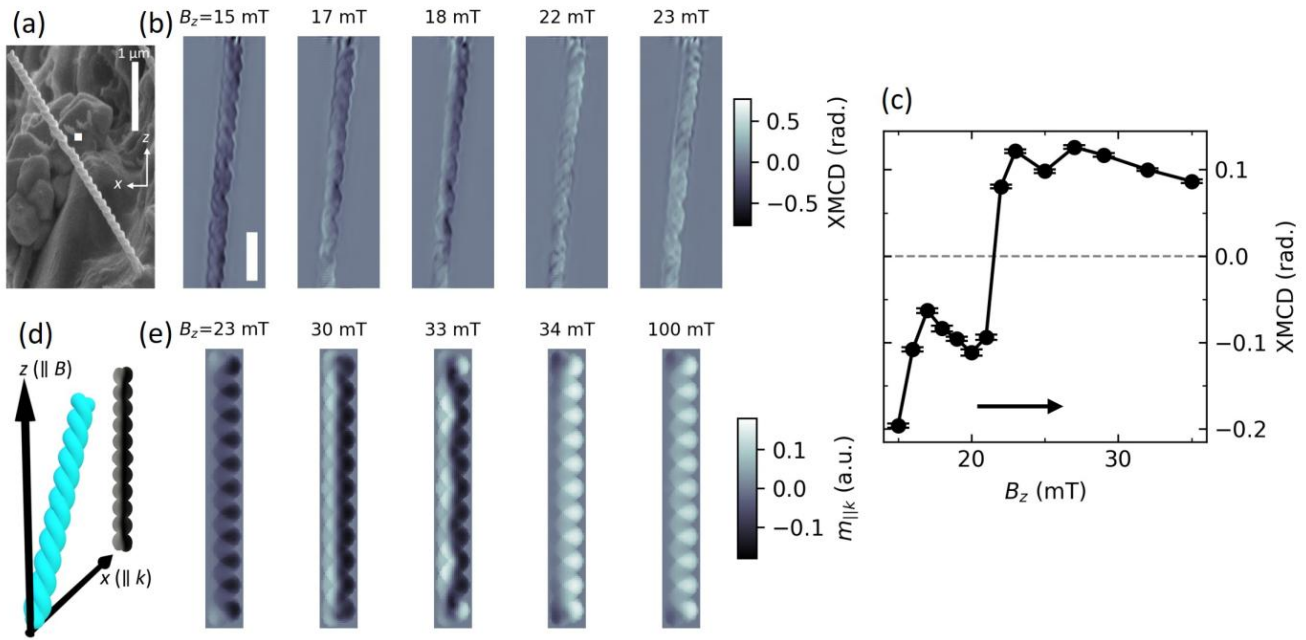

Fig. 4. (a) SEM image of a left-handed (LH) double helix nanowires fabricated on a pre-cut Cu TEM grid (image was taken at 45° stage tilt). A defect in the structure is marked by ▪. (b) Selected XMCD images taken during field sweeping from negative to positive direction of the z-axis. (c) Average XMCD signal as a function of the external magnetic field along the z-axis. Arrows also indicate the sweep direction, from negative to positive field. (d) Schematic illustration of the XMCD simulations. The structure is 30° tilted away from the vertical axis (z-axis), and an external magnetic field is applied along the z-axis. The XMCD signals were calculated as the dot product of the magnetization and the X-rays (parallel to the x-axis). (e) Calculated XMCD signal with the results of micromagnetic simulations. The magnetic field was applied along the z-axis, ranging from -100 to 100 mT in 1 mT increments.